\documentclass[mathleft
]{an}
\usepackage{graphicx}
\usepackage{times}
\usepackage{longtable}
\newcommand{\aap}{A\&A }

\newcommand{\nat}{Nature}
\newcommand{\memsai}{MmSAI}

\begin{document}

% The following seven commands are intended for editorial usage and should be ignored by
% the author(s).
\Pagespan{789}{}% Document's page range. 
% If second parameter is left empty, the last page is computed automatically.
\Yearpublication{2006}%
\Yearsubmission{2019}%
\Month{11}%   
\Volume{999}%  
\Issue{88}% 
% \DOI{This.is/not.aDOI}% 

\title{The rotation-activity relation of M dwarfs: From \textit{K2} to \textit{TESS} and \textit{PLATO}}

\author{St. Raetz\inst{1}\fnmsep\thanks{Corresponding author:
  \email{raetz@astro.uni-tuebingen.de}\newline}
%Example 
%for footnote, note the usage of the \texttt{fnmsep}
%command as separator between institute number and footnote mark} 
\and B. Stelzer\inst{1,2}
\and A. Scholz\inst{3}
}
%\titlerunning{Photometric analysis of 2MASS\,19090585+4911585}

   \institute{Institut f\"{u}r Astronomie und Astrophysik T\"{u}bingen (IAAT), Eberhard-Karls Universit\"{a}t T\"{u}bingen, Sand 1, D-72076 T\"{u}bingen, Germany\\
              \email{raetz@astro.uni-tuebingen.de}
         \and
             INAF - Osservatorio Astronomico di Palermo, Piazza del Parlamento 1, I-90134 Palermo, Italy
         \and
	    SUPA, School of Physics \& Astronomy, University of St. Andrews, North Haugh, St. Andrews KY 16 9SS, UK
}

\received{}
\accepted{}
\publonline{}


\abstract{Studies of the rotation-activity relation of late-type stars are essential to enhance our understanding of stellar dynamos and angular momentum evolution. We study the rotation-activity relation with \textit{K2} for M dwarfs where it is especially poorly understood. We analyzed the light curves of all bright and nearby M dwarfs form the Superblink proper motion catalog that were in the \textit{K2} field of view. For a sample of 430 M dwarfs observed in campaigns C0-C19 in long cadence mode we determined the rotation period and a wealth of activity diagnostics. Our study of the rotation-activity relation based on photometric activity indicators confirmed the previously published abrupt change of the activity level at a rotation period of $\sim$10\,d. Our more than three times larger sample increases the statistical significance of this finding.}

\maketitle

\section{Introduction}

Stellar activity is directly linked to magnetic fields that are believed to be generated and maintained by a dynamo which is driven by differential rotation and convection. Therefore, rotation and stellar activity are intimately connected.

The rotation-activity relation of late-type stars based on the measurements of the X-ray luminosity is empirically divided in two regimes: a ``saturated'' plateau of constant activity for fast rotators and a ``correlated'' regime for slow rotators. Because of the long spin-down times of M dwarfs compared to solar-like stars, rotation periods can become very long, depending on mass and age of the M dwarf (Scholz et al. 2013, Bouvier et al. 2014). Hence, for evolved, early M dwarfs, the long rotation periods and faint X-ray emission make the rotation-activity relation notoriously poorly defined.

By combining periods from the \textit{K2} mission with archival X-ray data Stelzer et al. (2016) started towards providing a statistical sample of bright and nearby M dwarfs with both known rotation period and X-ray detection. Photometric observations with space telescopes such as the \textit{K2} and \textit{\textit{TESS}} missions provide not only rotation periods even with low amplitudes but also a wealth of other activity diagnostics. The study of the rotation-activity relation based on photometric activity indicators by Stelzer et al. (2016) revealed, that, at a critical rotation period of $\sim$10\,d, the activity level changes abruptly. This drastic behavior is very different compared to the slowly decaying activity level seen in the correlated regime of the X-ray luminosity of M dwarfs (Wright, \& Drake 2016; Stelzer et al. 2016; Wright et al. 2018) or other activity indicators e.g. $H_{\alpha}$ (Newton et al. 2017). This phenomenon represents an open problem within the framework of dynamo theory.

NASA's \textit{K2} (\textit{Kepler} Two-Wheel) mission was a space telescope dedicated for optical photometric monitoring. It was the follow-up project of the main \textit{Kepler} mission after the failure of the second of four reaction wheels that were essential to maintain a very precise and stable pointing. The ``Second Light'' of \textit{Kepler}, the \textit{K2} mission which used a different observing concept, started its science observations in 2014 March (Howell et al. 2014). Until the end of the mission due to fuel exhaustion in 2018 October, \textit{K2} observed series of sequential observing "Campaigns" (campaign duration $\sim$80\,d) of fields distributed around the ecliptic plane. \textit{K2} provided continuous high precision light curves (LCs) in two cadence modes. The long cadence observations ($\sim$30\,min data point cadence) was the default observing mode while short cadence LCs ($\sim$1\,min data point cadence) were only provided for selected targets.

Stelzer et al. (2016) presented results for the relation between stellar rotation period and photometric activity indicators for a sample of 134 bright and nearby M dwarfs observed with the \textit{K2} mission in long cadence mode during campaigns C0-C4. Here we present the extension of this study to all bright and nearby M dwarfs from the Superblink proper motion catalog  (Lepine \& Gaidos 2011) in the \textit{K2} field of view until the end of the mission. With an increase of the sample size by a factor $>$3 we obtained better statistics to understand the behavior around the critical rotation period of $\sim$10\,d.

\section{The sample}

\begin{figure}
  \centering
  \includegraphics[width=0.32\textwidth,angle=270]{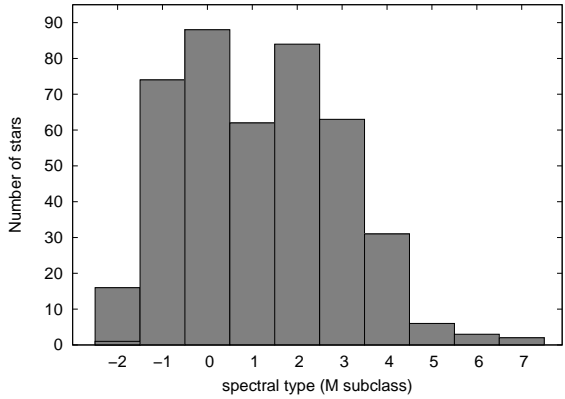}
  \caption{Distribution of the spectral types for our \textit{K2} M dwarf sample. The spectral types were obtained from the $V-J$ calibration given in Stelzer et al. (2016). Negative values denote spectral types earlier than M.}
  \label{SpT_Histogramm}
\end{figure}

Our study is based on the Superblink proper motion catalog  (Lepine \& Gaidos 2011), an all-sky catalog of $\sim$9000 M dwarf stars with apparent infrared magnitude $J< 10$. Until the end of the mission 430 targets during 20 \textit{K2} campaigns (C0-C19) were observed. Because of the overlap of some campaign fields we obtained in total 485 long cadence LCs.

\subsection{Stellar parameters}

The derivation of fundamental stellar parameters (effective temperature, mass, radius, and bolometric luminosity) was done as explained by Stelzer et al. (2016). In short, we used the empirical and semi-empirical relations of Mann et al. (2015; coefficients of the relations taken from the erratum, Mann et al. 2016) which are based on the colors $V-J$, $J-H$ and the absolute magnitude in the 2MASS $K$-band, $M_{\mathrm{Ks}}$. Magnitudes from the UCAC4 catalog (Zacharias et al. 2013) and the AAVSO Photometric All Sky Survey (APASS, Henden \& Munari 2014) were used for the calculations. To compute the $M_{\mathrm{Ks}}$ from the given magnitudes we used the empirical linear calibration of Stelzer et al. (2016, their Eq.~1). All stellar parameters and their uncertainties were then calculated using the following Monte Carlo approach. For each star we created 10\,000 data sets of $V-J$, $J-H$ and $M_{\mathrm{Ks}}$, randomly chosen from normal distributions defined by their error bars, and computed the stellar parameters. The final values were then determined as the mean and the standard deviation of the resulting distribution. The uncertainties of the relations of Mann et al. (2015) were included in the analysis. The distance was calculated from the distance modulus using the apparent magnitude in the 2MASS $K$-band and $M_{\mathrm{Ks}}$. Because of the higher precision of the $V$-band UCAC4 magnitude compared to the one provided by Lepine \& Gaidos (2011), we estimated a spectral type based on the $V-J$ color using the relations given in Stelzer et al. (2016, their Eq.~2 and 3). Our full target list with the stellar parameters is given in Table~\ref{stellar_param} in the appendix\footnote{The tables are available in ascii format at http://straetz.stiller-berg.de/7.html}, the distribution of the spectral types in our sample is shown in Fig.~\ref{SpT_Histogramm}.

\subsection{Multiplicity}
\label{multiplicity}

\begin{figure*}
  \centering
  \includegraphics[width=0.62\textwidth,angle=90]{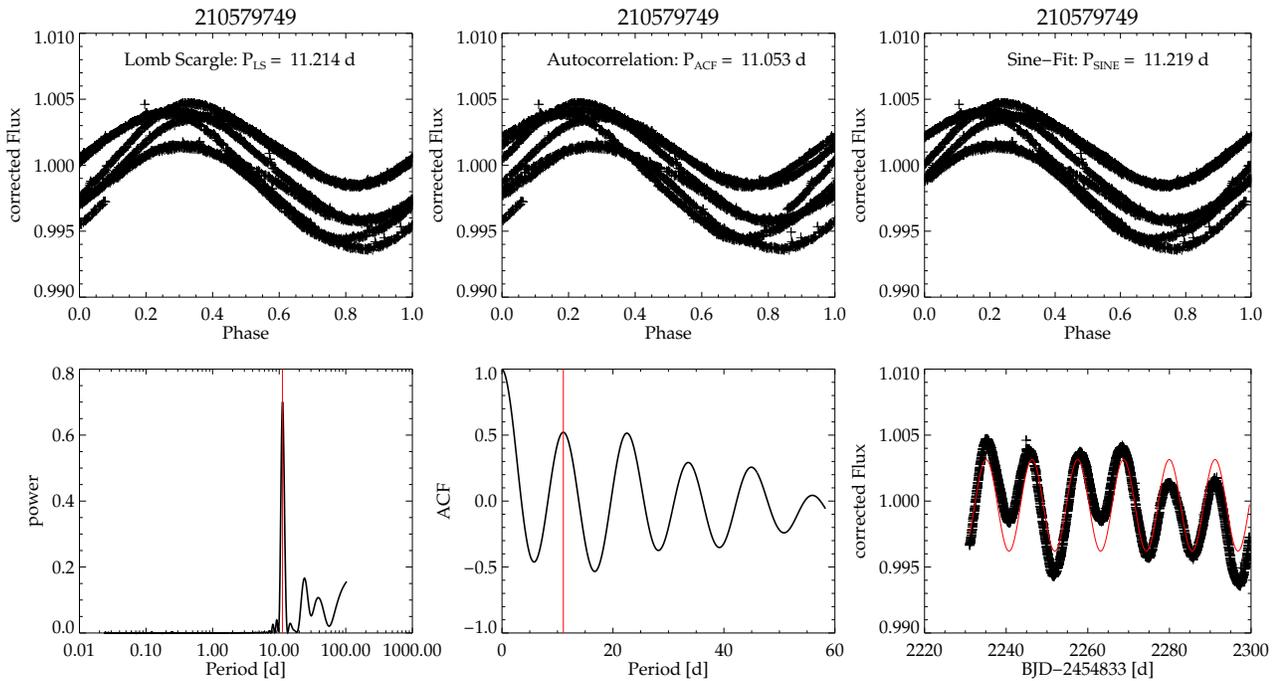}
  \caption{Example of the three period search methods LS, ACF and sine-fitting for EPIC\,210579749 observed in campaign C4. The top panels show the LCs phase-folded with the periods obtained with the different methods. The bottom panel shows the LS periodogram, the ACF and the original LC with the sine fit.}.
  \label{Phasfolded_LC_210579749}
\end{figure*}

Potential multiplicity of our target stars can affect our analysis in two ways. In the case of close visual binaries that are only separated by a few arcseconds the companion will contaminate the photometry of our target. On the other hand, very close eclipsing or spectroscopic binaries are well known to show strong flares and an enhanced flare frequency as shown in studies by, e.g., Mathioudakis et al. (1992),  Morgan et al. (2012), or  Skinner et al. (2017). Hence, knowing the multiplicity status of our stars is crucial for the interpretation of our results. Therefore, we searched all stars in our \textit{K2} M dwarf sample for evidence of multiplicity.

One prominent example from our sample is the common proper motion pair CU\,Cnc (EPIC\,211944670, GJ\,2069A, HIP\,41824) and CV\,Cnc (EPIC\,211944856 ,GJ 2069B) that is actually a quintuple system. CU\,Cnc is a known eclipsing M dwarf binary (Delfosse et al. 1999a) with a direct imaging companion candidate (Beuzit et al. 2004) while CV\,Cnc is itself a visual binary (Delfosse et al. 1999b). In Raetz et al. (2020) we showed in a systematic study comparing rotation and activity in short- and long-cadence \textit{K2} data of M dwarfs, that the highest flare rate in our sample was observed on CU\,Cnc. This strengthens the hypothesis that flares might be induced by the presence of a close companion.

We searched all 430 stars in the Washington Visual Double Star (WDS) Catalog (Mason et al. 2001) for information about multiplicity. In order to check for eclipsing and spectroscopic binaries we matched the General Catalogue of Variable Stars (Samus et al. 2017) and the 9th Catalogue of Spectroscopic Binary Orbits (Pourbaix et al. 2004) to our target list. To search for additional visual binaries we conducted a common proper motion analysis for all targets in our sample using the \textit{Gaia} Data Release 2 (Gaia Collaboration et al. 2018). We downloaded all \textit{Gaia} sources within a radius of 1\,arcmin ($\sim$ size of a \textit{K2} target pixel file) around each of the M dwarfs. Then we compared for each of these lists the proper motions in right ascension and declination of all Gaia sources to the proper motion of our M dwarf target. If the proper motion of a \textit{Gaia} source was within 10\% of the proper motion of the M dwarf we considered them as a proper motion pair. We found 38 pairs in 34 systems (for four pairs both components have an individual EPIC number and both belong to our M dwarf sample) of which 26 were also listed in the WDS catalog.

Stelzer et al. (2016) list 15 targets from our sample that show evidence for binarity which are included in our compilation. In total, we found evidence of multiplicity for 96 targets. They are all given in Table~\ref{binaryTable} in the appendix. The visual companions from the WDS catalog have separations between 0.1 and 305\,arcsec. Companions with separations larger than $\sim$40\,arcsec (10 pixels of \textit{K2} with an image scale of 3.98\,arcsec per pixel) will not influence our analysis and are considered as single stars in the following. Consequently, 30 out of the 96 targets with evidence for multiplicity were subsequently classified as singles. The fraction of stars from the whole sample with evidences for companions is $\sim$15\%. That is slightly higher but similar to the results of Stelzer et al. (2016, $\sim$11\% binarity fraction.)

\section{\textit{K2} Data analysis}

\begin{figure}
  \centering
  \includegraphics[width=0.32\textwidth,angle=270]{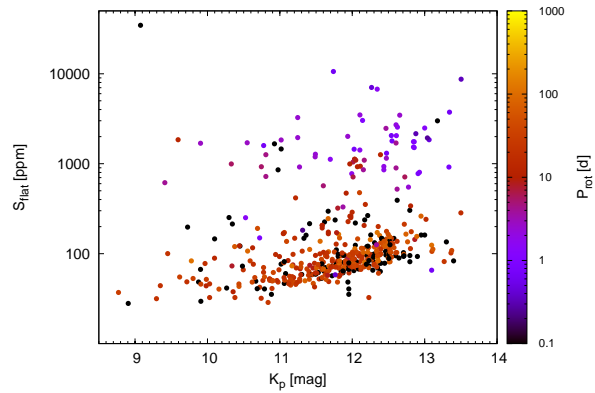}
  \caption{Standard deviation of the flattened and cleaned LC $S_{\mathrm{flat}}$ over magnitude in the \textit{Kepler} band $K_{\rm p}$. Color-coded is the rotation period of the stars. The stars were we could not detect any period are shown in black.}
  \label{Kp_SFlat}
\end{figure}

\begin{figure}
  \centering
  \includegraphics[width=0.45\textwidth]{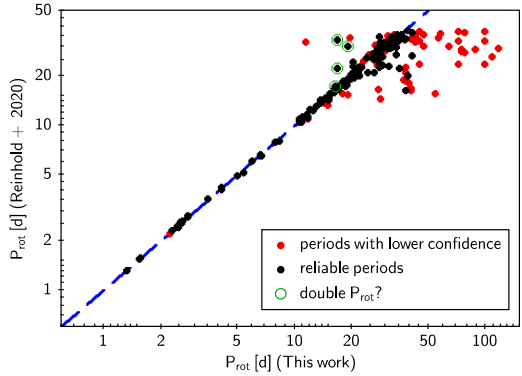}
  \caption{Comparison between the period measurements from this work with the ones from Reinhold et al. (2020). Rotation periods of 161 stars are found in both samples. The black and red symbols denote the stars that we flagged as having reliable periods and periods with lower confidence, respectively. The green circles mark the stars where the LC is ambiguous between a period and twice its value.}
  \label{periods_comp_Reinhold}
\end{figure}

\begin{figure}
  \centering
  \includegraphics[width=0.32\textwidth,angle=270]{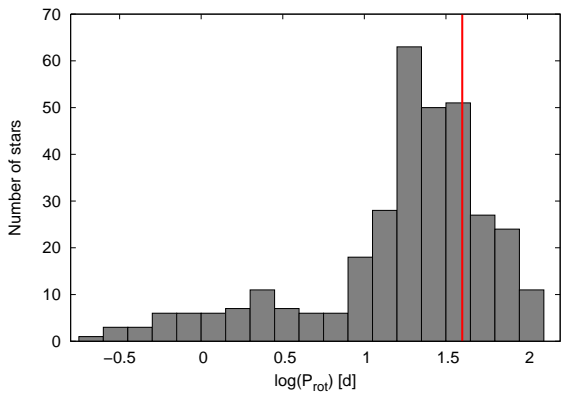}
  \caption{Distribution of $P_{\rm rot}$ for the 341 M dwarfs with a measurable rotation period. The red solid line marks our defined transition between reliable periods and periods with less confidence.}
  \label{period_Histogramm}
\end{figure}

Vanderburg \& Johnson (2014) provide fully reduced and detrended LCs for all targets from our \textit{K2} M dwarf sample that we downloaded from the website of A. Vanderburg\footnote{https://www.cfa.harvard.edu/$\sim$avanderb/k2.html}.

For our analysis of the rotation-activity relation we extracted the following parameters from each individual LC: (1) rotation period, (2) photometric activity indicators (standard deviation of the full LC, standard deviation of the flattened and cleaned LC, amplitude of the rotational signal, flare frequency, peak flare amplitude). The definition of these properties are given in Sect.~\ref{activity1}, \ref{activity2} and \ref{activity3}. A detailed description of the determination of these properties can be found in Stelzer et al. (2016). The rotation periods and the values for the measured photometric activity indicators for our targets are given in Table~\ref{periodTable} in the appendix.

\subsection{Period search}

The rotation period, $P_{\rm rot}$, was measured with standard time series analysis techniques, a Lomb Scargle periodogram (LS; Lomb 1976 and Scargle 1982) and the autocorrelation function (ACF). Finally as a third method, we fitted the LCs with a sine function which also yields a tentative period estimate for stars that exceed the K2 monitoring baseline of $\sim$70-80\,d. The LCs of our sample were phase-folded with the periods of all three methods. In Fig.~\ref{Phasfolded_LC_210579749} an example for the different period search methods for one of the targets from our sample is shown. A file with such figures for all 485 LCs from the 430 targets in our \textit{K2} M dwarf sample is available online. Among the results obtained for the three different methods, the period that best represents the LCs was selected by eye-inspection and was then used in the further analysis. Uncertainties on the rotation periods are given as the standard deviation of the period estimates from the different methods.

\subsection{Activity diagnostics from rotation cycles}
\label{activity1}

Because $P_{\rm rot}$ is derived from the periodic brightness variations that are caused by cool spots on the stellar surface, the period search yields simultaneously the amplitude of photometric variability associated with these spots. The amplitude is, hence, an indication of the stellar activity level. We determined the full amplitude $R_{\mathrm{per}}$ (given in \%) as the mean value of all amplitudes measured for each rotation cycle as defined by McQuillan et al. (2013).

In addition to $R_{\mathrm{per}}$ we also determined the standard deviation of the full LC $S_{\mathrm{ph}}$ which includes the rotational signal and all outliers. $S_{\mathrm{ph}}$ was introduced by Mathur et al. (2014) as a measure of the global evolution of the variability and is, hence, an important activity indicator.

\subsection{Flare identification and validation}
\label{activity2}

The search for flares in the K2 LCs was done with an iterative process which consists of three steps. Firstly, boxcar smoothing of the original LC was applied. Then the smoothed LC was subtracted from the original LC which resulted in a flat LC with the rotational signal removed. Finally, all outliers that deviated by more than 3$\sigma$ were flagged and removed. These 3 steps were repeated three times with decreasing width of the boxcar. The initial width of the boxcar was selected individually based on the measured rotation period. The steps of decreasing width were chosen by testing different combinations and checking the resulting flattened LC. The iterative process results in two types of LCs: (1) a flattened LC that includes all flagged outlying points, (2) a flattened and cleaned LC with both, the rotational signal and the outliers removed.

The flattened LC was used as basis for the flare identification. From all the flagged outlying points only groups of at least two consecutive upwards outliers were selected as flare candidates. To validated these candidates as flares an additional criterion was introduced as an indication of the flare shape. We require that the ratio between the flux of the flare maximum is two times higher that the flux of the last flare point (the last consecutive outlying point) which also implies, that the maximum was not allowed to be the last flare point.

For each validated flare we determined the flare amplitude, $A_{\mathrm{peak}}$, by subtracting the continuum flux level given by the flattened LC at the time of the flare peak from the corresponding flux.

\subsection{Noise level and detection bias}
\label{activity3}

\begin{figure*}
  \centering
  \includegraphics[width=0.32\textwidth,angle=270]{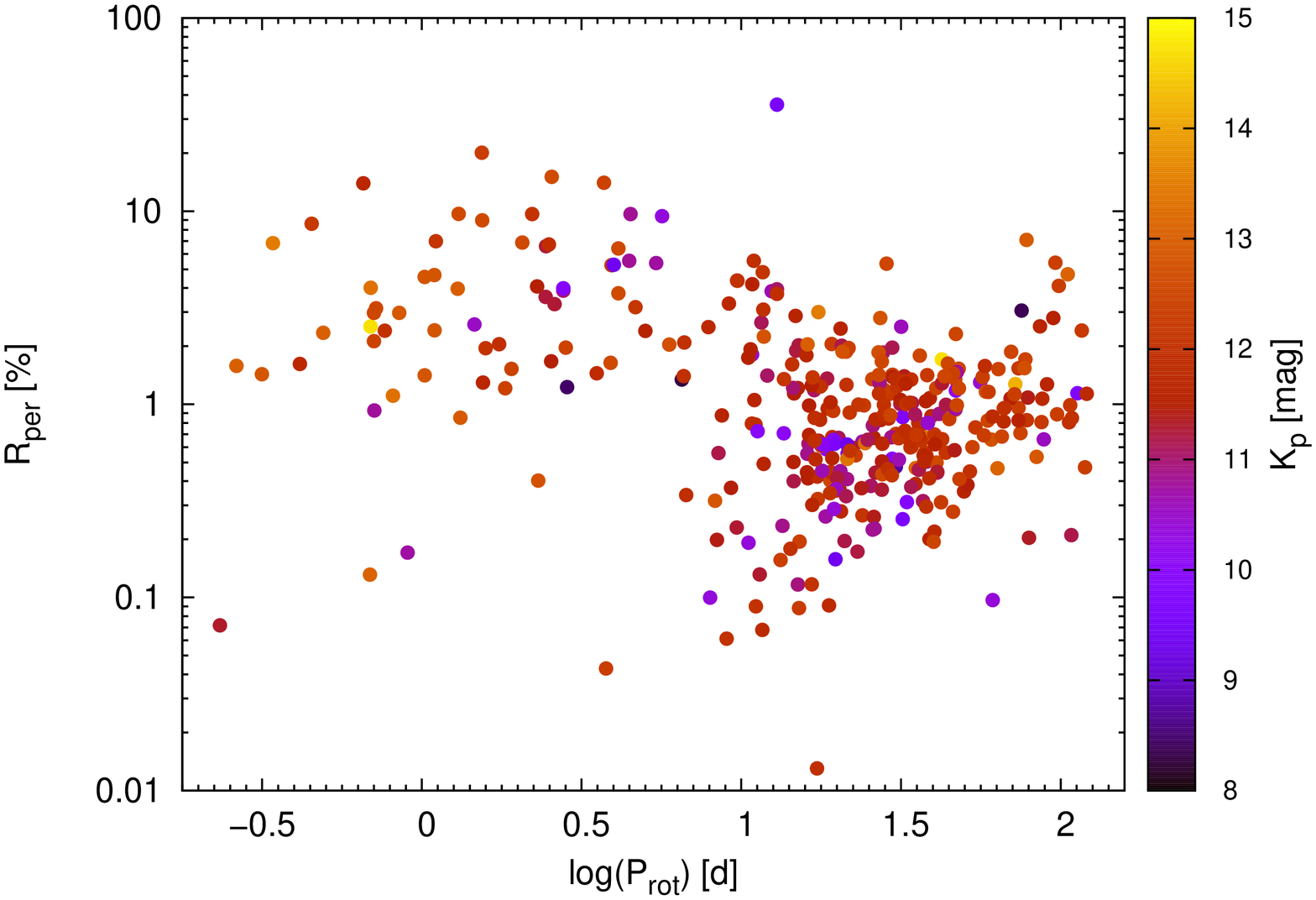}
  \includegraphics[width=0.32\textwidth,angle=270]{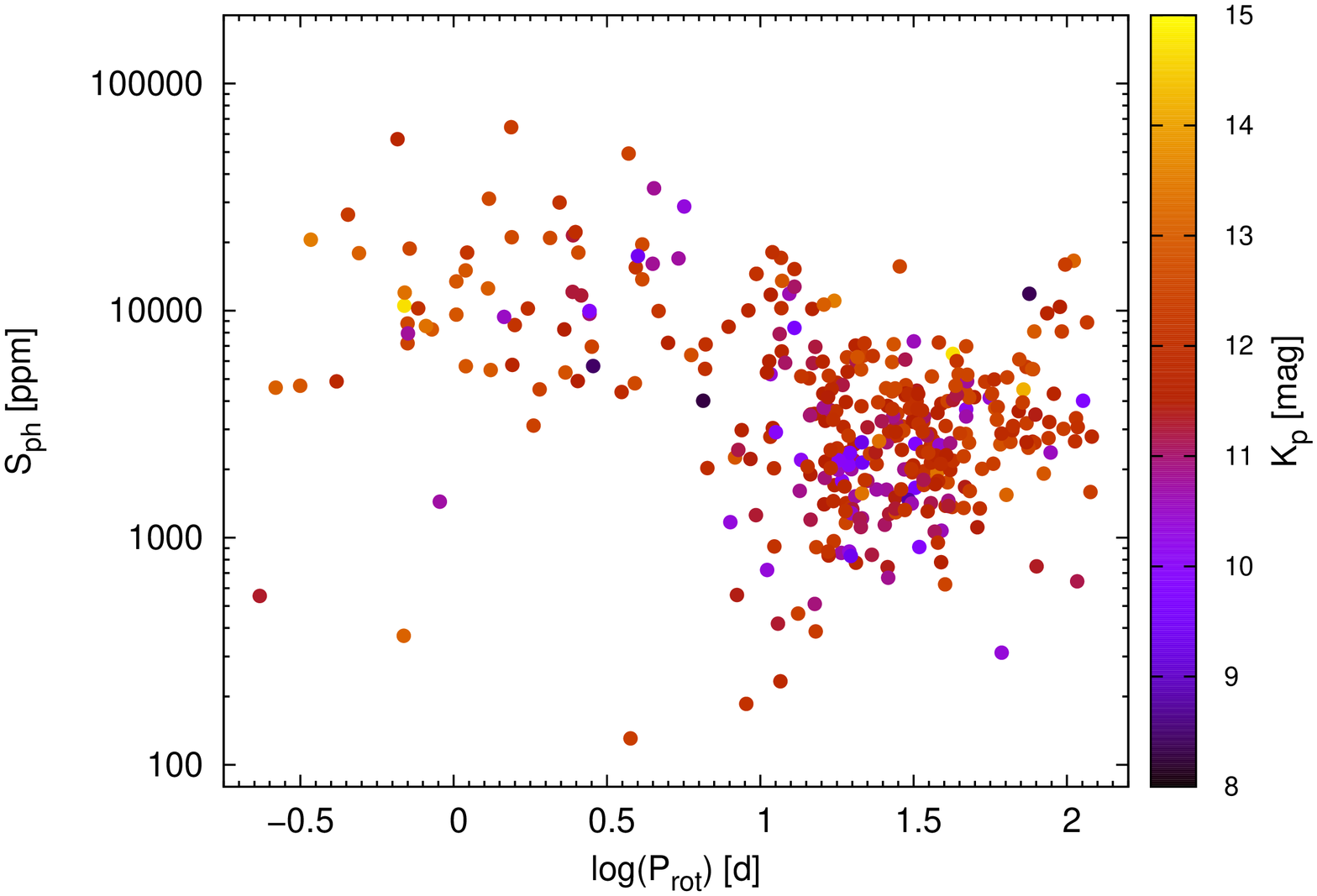}
  \includegraphics[width=0.32\textwidth,angle=270]{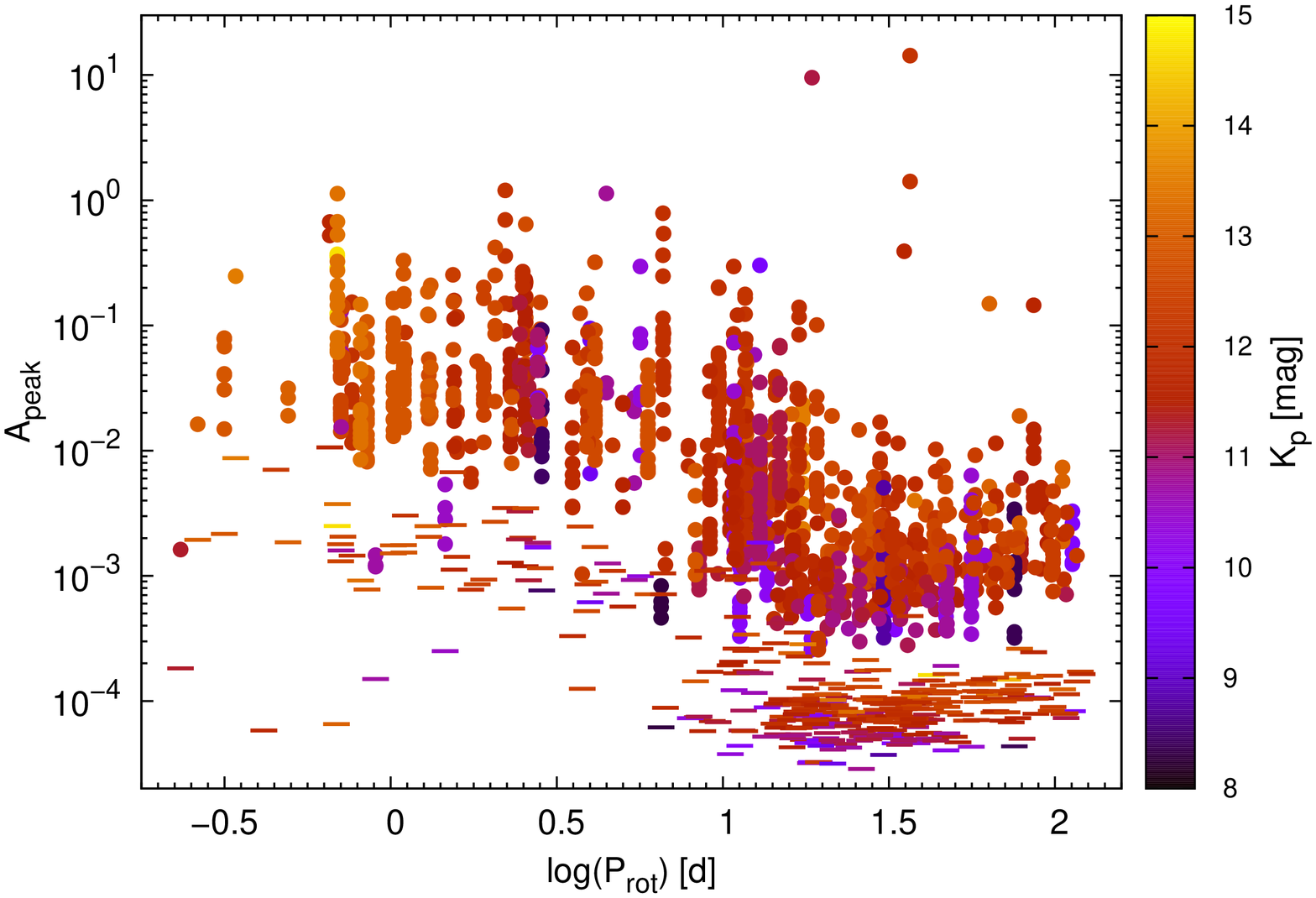}
  \includegraphics[width=0.32\textwidth,angle=270]{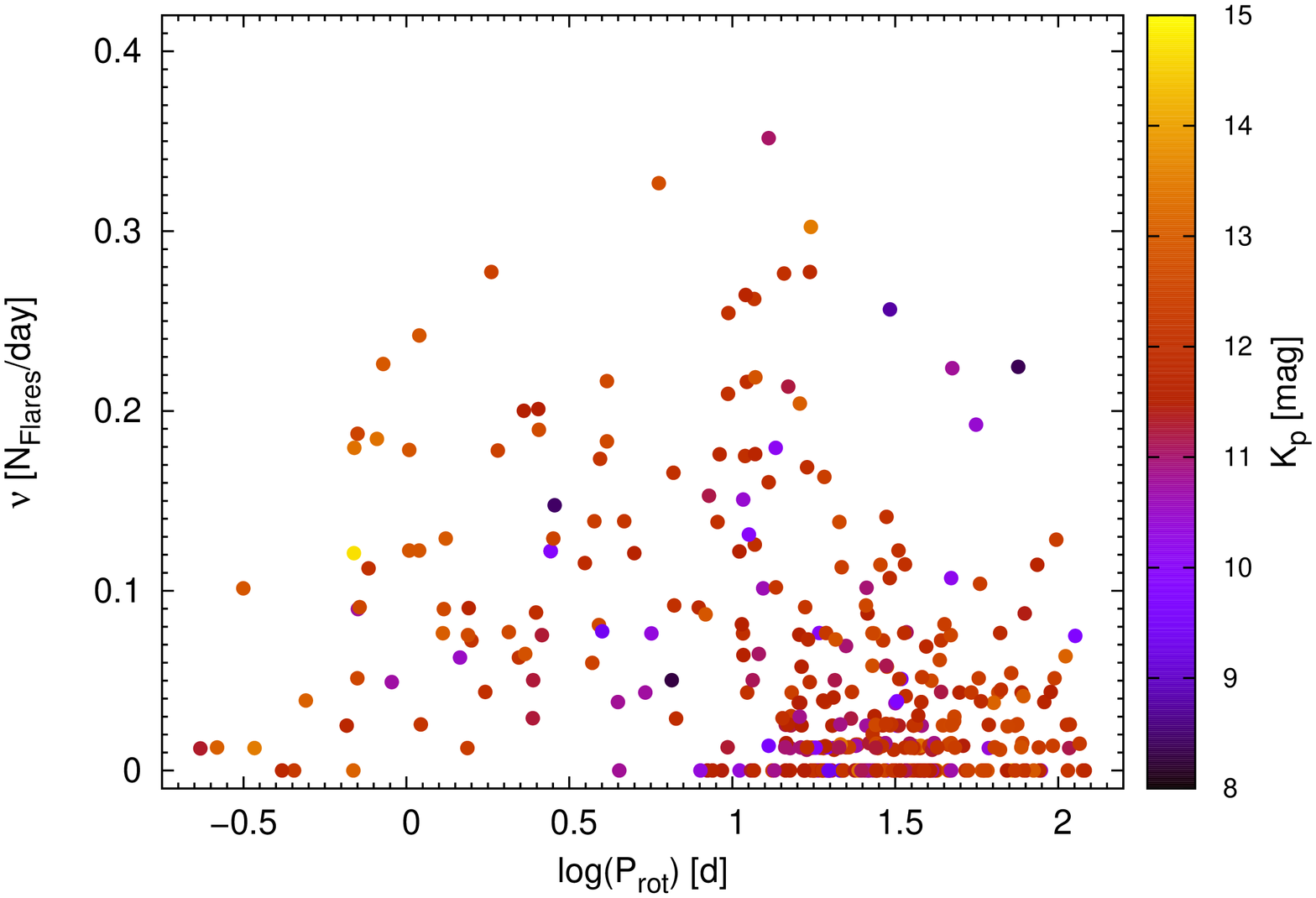}  
  \caption{Relation between rotation period and various activity indicators for all 341 targets with measurable rotation period. \textbf{Top left:} full amplitude, $R_{\mathrm{per}}$, of the rotational signal given in \%. \textbf{Top right:} standard deviation of the full LC, $S_{\mathrm{ph}}$. \textbf{Bottom left:} flare amplitude, $A_{\mathrm{peak}}$, versus rotation period for all targets that show flares. The small horizontal bars represent the standard deviation of the cleaned and flattened LCs ($S_{\mathrm{flat}}$) which is a measure of our detection threshold. \textbf{Bottom right:} flare frequency, $\nu$.}
  \label{period_activity}
\end{figure*}

To investigate our flare detection bias we determined the standard deviation of the flattened and cleaned LC, $S_{\mathrm{flat}}$. This value can be seen as a representation of the noise level in our data. In Fig.~\ref{Kp_SFlat} we show the $S_{\mathrm{flat}}$ over the Kepler magnitude $K_{\rm p}$. As expected the photometric noise increases towards fainter stars. In addition to the magnitude dependence of $S_{\mathrm{flat}}$ we see an additional scatter for a given $K_{\rm p}$. Stelzer et al. (2016) found that fast rotators ($P_{\rm rot} < 10$) have a higher $S_{\mathrm{flat}}$ than the slow rotators ($P_{\rm rot} > 10$\,d) and concluded that there is an additional underlying noise of astrophysical origin e.g. unresolved mini-flares or mini-spots for the fast rotating stars. Hence, the large scatter of  $S_{\mathrm{flat}}$ for a given $K_{\rm p}$ is an indication that our sample comprises targets of various rotation periods. This is confirmed by the color-code in Fig.~\ref{Kp_SFlat}. The fast rotators are located well above the group of the slow rotators. With our more than three times larger sample than the one of Stelzer et al. (2016) we are able, for the first time, to distinguish two M dwarf populations in $K_{\rm p}$ - $S_{\mathrm{flat}}$ space.

\section{Results}

\begin{figure}
  \centering
  \includegraphics[width=0.32\textwidth,angle=270]{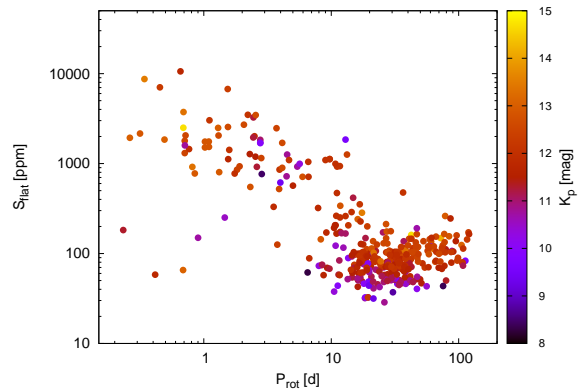}
  \caption{Standard deviation of the flattened and cleaned LC, $S_{\mathrm{flat}}$, versus rotation period for the 341 stars with measurable rotation period.}
  \label{period_SFlat_Kp}
\end{figure}

After the eye-inspection of the phase-folded LCs resulting from the period search we sorted the targets in three different groups: those with reliable periods (flag `Y' in Table~\ref{periodTable}), periods with lower confidence (flag `?') and no detected period (flag `N'). Since a reliable period estimate cannot be obtained with less than two full cycles we considered all periods greater than $\sim$40\,d (depending on the duration of the individual campaigns) as periods with lower confidence. We found 201, 140, and 89 stars to be in the groups  reliable periods,  periods with lower confidence, and no period, respectively. Example LCs for these three groups are shown in Fig.~\ref{Bsp_LC_K2} in the appendix. For several stars all period search methods show hints of a period that is a multiple of the determined value. These stars are flagged in Table~\ref{periodTable}. For EPIC\,210579749 that is shown in Fig.~\ref{Phasfolded_LC_210579749}, the highest peak in the LS periodogram is $\sim$4 times larger than the one of the double period. The value corresponding to our most significant period, ~11d, was also found by Armstrong et al. (2016). In total, we found rotation periods for 341 stars, which is $\sim$79\% of all targets in our M dwarf sample. This value is in excellent agreement with the study of Stelzer et al. (2016) who already presented the rotation periods for the  Lepine \& Gaidos (2011) M dwarfs from campaigns C0-C4. Recently, Reinhold et al. (2020) published rotation periods for stars observed in \textit{K2} campaigns C0-C18. We found that 161 stars from our sample have period measurements in Reinhold et al. (2020). The result of our rotation period comparison is shown in Fig.~\ref{periods_comp_Reinhold}. For periods up to $\sim$25\,d our measurements are in excellent agreement. Most of the deviations in the longer period regime are stars with periods with lower confidence where we also expect to measure fractions or multiples of the true rotation periods. About 78\% of our targets have $P_{\rm rot} > 10$\,d, and $\sim$10\% have $P_{\rm rot} > 65$\,d (see the distribution of $P_{\rm rot}$ in Fig.~\ref{period_Histogramm}).

In Fig.~\ref{period_activity} we show our measured activity diagnostics versus the rotation period. The upper panels show the full amplitude of the rotation signal, $R_{\mathrm{per}}$, and the standard deviation of the full LC, $S_{\mathrm{ph}}$. Both parameters show a clear trend to higher values for faster rotators. In particular we found larger spot amplitudes for stars with shorter rotation periods. This finding is in good agreement with the measurements of McQuillan et al.(2013) and Reinhold et al. (2020). The bottom panel of Fig.~\ref{period_activity} show the parameters connected to flares for all stars with detected rotation period that show flares (272 stars). Again, we see a change of behavior at $P_{\rm rot} \sim 10$\,d. The faster rotators tend to have more flares (a higher flare rate) and the flares tend to have larger relative amplitudes. The larger spot amplitudes, the higher number of flares and the larger flare amplitudes account for the larger value of $S_{\mathrm{ph}}$ in the fast rotator regime.

All four plots in Fig.~\ref{period_activity} are color-coded according to the \textit{Kepler} magnitude to check if our results are biased because of different detection sensitivities. We found no evidence for such a bias since the bright and faint stars are evenly distributed among the fast and slow rotators. 

The small horizontal bars in the rotation period - peak flare amplitude plot (bottom left plot of Fig.~\ref{period_activity}) represent the standard deviation of the cleaned and flattened LCs ($S_{\mathrm{flat}}$) which is a measure of our detection threshold. Although we are not able to detect the smallest flares for the fast rotators, the difference in the upper envelope is a real effect. This upper envelope is defined by the maximum flare amplitudes $A_{\mathrm{peak,max}}$ of the stars. We have calculated the mean of the $A_{\mathrm{peak,max}}$ values separately for fast and slow rotators, and it is $\sim$2 times higher for the fast rotators than for the slow rotators.

The flare frequency, $\nu$, versus the rotation period is shown in the bottom right plot of Fig.~\ref{period_activity}. For many stars with rotation periods $P_{\rm rot} > 10$\,d we detected none or only one flare and, hence, $\nu$ is close to zero $N_{\mathrm{flares}}/d$. In the fast rotator regime, $P_{\rm rot} < 10$\,d, we barely found LCs without any flares. With an average flare rate of 0.10\,$N_{\mathrm{flares}}/d$ and 0.04\,$N_{\mathrm{flares}}/d$ for the fast and slow rotators, respectively, we obtained $\nu_{P_{\rm rot} < 10\rm d}  \sim2.5\times\nu_{P_{\rm rot} > 10\rm d}$. We determined the average flare frequency for the whole sample to be  $\nu = 0.054\, N_{\mathrm{flares}}/d$. Note, that this number strongly depends on how many fast and slow rotators are included in our sample. Since our sample consists of a high number of slow rotators the average $\nu$ is biased towards a smaller value.

To examine if the multiplicity of our targets affect the flare parameters we splitted our sample of stars with measured rotation period in multiple systems and single stars. Then we computed for both groups the mean values for $\nu$ for all, only short and only long $P_{\rm rot}$, and the maximum peak flare amplitude. We found that all values are higher for the multiple systems. To check if these results are significant, we ran a Student's T-test. The results are summarized in Table~\ref{T-test-results}. All p-values suggest that the mean values of $\nu$ for most of the tested groups are equal for singles and multiple systems. Only in the group of the complete sample we see a slight statistically significant difference. Hence, the multiple systems in our sample only marginally affect our results on the rotation-activity relation in terms of flare parameters. In Fig.~\ref{period_activity_multiplicity} in the appendix we re-plotted the relation between rotation period and flare parameters (bottom panels of Fig.~\ref{period_activity}) with the multiple systems marked.

\begin{figure*}
  \centering
  \includegraphics[width=0.3\textwidth,angle=270]{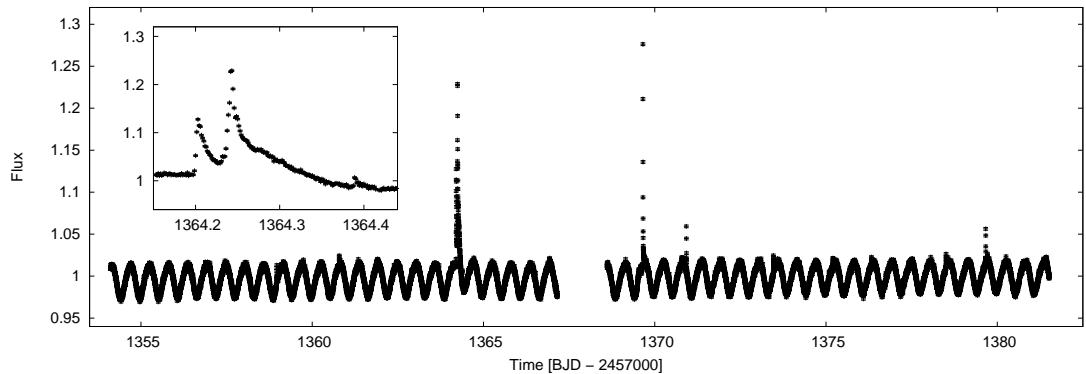}
  \caption{Example of a Lepine \& Gaidos (2011) M2 dwarf observed in \textit{TESS} sector 2 with a clear rotation signal and a number of strong flares. The small plot shows a zoom on the first flare which consists of multiple peaks.}
  \label{TIC471015484_norm_black}
\end{figure*}

\begin{table}[]
\centering
\caption{Results of the T-test for the comparison of flare parameters between single and multiple stars. $A_{\mathrm{peak,max}}$ is the maximum peak flare amplitude, $\overline{\nu}$ is the average flare frequency evaluated for three samples: all stars, only fast and only slow rotators. }
\label{T-test-results}
\begin{tabular}{lcccc}
\hline \hline
Parameter & singles & multiples & T-statistic & p-value\\\hline
$A_{\mathrm{peak,max}}$ & 9.5 & 14.3 & & \\
$\overline{\nu}$ & 0.051 & 0.074 & -2.4 & 0.016 \\
$\overline{\nu}_{P_{\rm rot}<10\,\mathrm{d}}$ & 0.097 & 0.118 & -1.2 & 0.251 \\
$\overline{\nu}_{P_{\rm rot}\geq10\,\mathrm{d}}$ & 0.040 & 0.051 & -1.1 & 0.267 \\
\hline
\end{tabular}
\end{table}

Finally, we examine the behavior $S_{\mathrm{flat}}$ with the rotation period. The result is shown in Fig.~\ref{period_SFlat_Kp}. As already noticed by Stelzer et al. (2016) in the more than three times smaller sample from campaigns C0-C4, we see a clear trend of higher $S_{\mathrm{flat}}$ values for lower $P_{\rm rot}$. Since the same trend that is visible in the activity diagnostics related to the rotation cycle and flares this might be an indication that this additional noise is caused by unresolved spot and flare activity. Hence, the underlying noise is most likely of astrophysical origin.  

To summarize, we found a bimodal distribution with fast rotators showing a higher activity level than slow rotators for all examined activity diagnostics. The critical rotation period where the activity behavior changes is $\sim$10\,d. With a sample that is $>$3 times larger, we could confirm the findings by Stelzer et al. (2016). We did not find statistically significant evidence, that the multiplicity status of our targets influence our final results.

\section{Outlook}

Despite the outstanding capabilities of K2, the short observational baseline  does not allow us to explore the full range of M dwarf rotation periods and the low cadence limits our study to long flares with durations of $>$1\,h. Only 56 of the 430 targets in our \textit{K2} M dwarf sample were observed by \textit{K2} in short cadence mode. The comparison of the long and short cadence data are presented in detail by Raetz et al. (2020). Our study of the rotation-activity relation of M dwarfs will strongly benefit from the observation with other existing and upcoming photometric space missions i.e. \textit{TESS} and \textit{PLATO}.

NASA's \textit{TESS} (Transiting Exoplanet Survey Satellite) mission is an optical space telescope dedicated for the search for planets transiting nearby stars. \textit{TESS} was launched on April 18, 2018 and performs a near all-sky survey. The brightest $\sim$200\,000 stars that are summarized in the Candidate Target List (CTL, Stassun et al. 2019) are observed in a 2-min cadence mode. In its two years prime mission (the mission was already extended) $\sim$7000 M dwarfs from the Lepine \& Gaidos (2011) catalog are observed. All of these M dwarfs are bright and will be included in the CTL. Hence, all targets will be observed with 2-min cadence. \textit{TESS} targets during its prime mission 26 different sectors of the sky. Due to the overlap of the sectors \textit{TESS} delivers $\sim$11\,500 LCs in two years of operation. For about 5000 Lepine \& Gaidos (2011) M dwarfs \textit{TESS} will provide single LCs with an observational baseline of $\sim$27d. However, $\sim$1\% of the targets are in the so-called continuous viewing zone where \textit{TESS} provides thirteen 27d-light curves. Although for most of the targets only short rotation periods can be found \textit{TESS} provides a excellent coverage for a subsample of our targets. An example of a bright ($T=10.8$\,mag) M2 dwarf observed in \textit{TESS} sector 2 with a clear rotation signal and a number of strong flares is shown in Fig.~\ref{TIC471015484_norm_black}. \textit{TESS} drastically enlarges our sample of bright M dwarfs continuously monitored at high cadence and allows us to study in depth the activity and in particular the morphology of flares due to the 2-min cadence. 

\begin{figure}
  \centering
  \includegraphics[width=0.3\textwidth,angle=270]{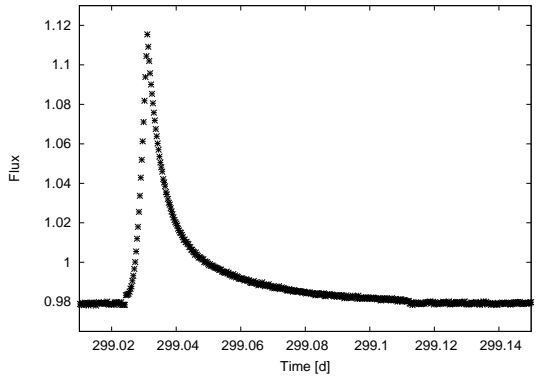}
  \caption{Example of a flare on a simulated \textit{PLATO} LC. The flare was constructed using the template by Davenport et al. (2014)}
  \label{Plato_LC_Flares_zoom_black}
\end{figure}

PLAnetary Transits and Oscillations of stars (\textit{PLATO}) is a M-class mission in ESA's Cosmic Vision programme and is foreseen to be launched in 2026. Its main goal is to find and study a large number of extrasolar planetary systems. As a photometric mission it will provide continuous high precision observations for at least 300\,000 stars with a very short cadence of 25\,s. This cadence will enable detailed studies of the flare shape. An example for a flare on a simulated \textit{PLATO} LC is shown in Fig.~\ref{Plato_LC_Flares_zoom_black}. The \textit{PLATO} fields, that are not finally decided yet, will likely include 2 long fields with an observing time of up to two years each and an additional step-and stare phase with durations of several months. If we consider the probable fields given in Barban et al. (2013), \textit{PLATO} will observe $\sim$3500 Lepine \& Gaidos (2011) M dwarfs, $\sim$500 of them in the two long fields. \textit{PLATO} with its unprecedented precision, short cadence and long observational baseline, will allow us to study the magnetic activity indicators in up to now unrivaled detail. 

\acknowledgements
We would especially like to thank A. Vanderburg for his public release of the analyzed K2 light curves, upon which much of the present work is based.

%\newpage

\appendix

\section{M dwarf sample: stellar parameters and rotation-activity measurements}

\onecolumn
\begin{longtable}[h]{lccr@{\,$\pm$\,}lr@{\,$\pm$\,}lr@{\,$\pm$\,}lr@{\,$\pm$\,}lr@{\,$\pm$\,}lr@{\,$\pm$\,}lc}
\caption{Stellar parameters of our \textit{K2} M dwarf sample. Fourty-eight targets were observed in multiple campaigns.}
\label{stellar_param}
\\ \hline\hline
EPIC ID & Camp. & $K_{\rm p}$ & \multicolumn{2}{c}{$M_{\mathrm{Ks}}$} & \multicolumn{2}{c}{$R_{\mathrm{\ast}}$  } & \multicolumn{2}{c}{ $M_{\mathrm{\ast}}$ } &  \multicolumn{2}{c}{$T_{\mathrm{eff}}$} & \multicolumn{2}{c}{log\,$L_{\rm bol}$} & \multicolumn{2}{c}{d} & SpT \\ 
&  &  & \multicolumn{2}{c}{mag} & \multicolumn{2}{c}{[R$_{\odot}$]} & \multicolumn{2}{c}{ [M$_{\odot}$]} &  \multicolumn{2}{c}{[K]} & \multicolumn{2}{c}{[L$_{\odot}$]} & \multicolumn{2}{c}{[pc]} &  \\\hline
\endfirsthead
\multicolumn{16}{l}{Table~\ref{stellar_param} continued.}\\
\hline\hline
EPIC ID & Camp. & $K_{\rm p}$ & \multicolumn{2}{c}{$M_{\mathrm{Ks}}$} & \multicolumn{2}{c}{$R_{\mathrm{\ast}}$  } & \multicolumn{2}{c}{ $M_{\mathrm{\ast}}$ } &  \multicolumn{2}{c}{$T_{\mathrm{eff}}$} & \multicolumn{2}{c}{log\,$L_{\rm bol}$} & \multicolumn{2}{c}{d} & SpT \\ 
&  &  & \multicolumn{2}{c}{mag} & \multicolumn{2}{c}{[R$_{\odot}$]} & \multicolumn{2}{c}{ [M$_{\odot}$]} &  \multicolumn{2}{c}{[K]} & \multicolumn{2}{c}{[L$_{\odot}$]} & \multicolumn{2}{c}{[pc]} &  \\\hline
\endhead
\hline
\multicolumn{16}{c}{to be continued on the next page}\\
\hline
\endfoot
\endlastfoot
202059183	&	C0	&	13.30	&	5.94	&	0.12	&	0.45	&	0.02	&	0.47	&	0.02	&	3503	&	84	&	-1.53	&	0.05	&	41.0	&	2.2	&	M2.5	\\
202059188	&	C0	&	14.70	&	7.33	&	0.09	&	0.27	&	0.01	&	0.27	&	0.01	&	3179	&	81	&	-2.14	&	0.04	&	22.1	&	0.9	&	M4.5	\\
202059190	&	C0	&	12.90	&	4.40	&	0.04	&	0.73	&	0.02	&	0.73	&	0.01	&	4069	&	81	&	-0.83	&	0.02	&	90.7	&	2.0	&	K7.0	\\
202059192	&	C0	&	13.10	&	5.26	&	0.05	&	0.56	&	0.02	&	0.59	&	0.01	&	3713	&	81	&	-1.23	&	0.03	&	60.1	&	1.5	&	M1.0	\\
202059193	&	C0	&	12.50	&	4.93	&	0.06	&	0.63	&	0.02	&	0.65	&	0.02	&	3834	&	82	&	-1.08	&	0.03	&	57.1	&	1.7	&	M0.0	\\
202059195	&	C0	&	14.70	&	7.00	&	0.10	&	0.31	&	0.02	&	0.31	&	0.01	&	3312	&	82	&	-1.99	&	0.04	&	26.0	&	1.2	&	M4.0	\\
202059197	&	C0	&	12.30	&	4.64	&	0.04	&	0.68	&	0.02	&	0.69	&	0.01	&	3964	&	81	&	-0.95	&	0.02	&	64.1	&	1.4	&	M0.0	\\
202059198	&	C0	&	11.60	&	5.31	&	0.10	&	0.56	&	0.02	&	0.58	&	0.02	&	3727	&	99	&	-1.25	&	0.04	&	29.3	&	1.4	&	M1.0	\\
\end{longtable}

\begin{longtable}[h]{ccr@{\,$\pm$\,}lccr@{\,$\pm$\,}lcccc}
\caption{Rotation period and activity indicators for all 430 targets. For the 48 targets that were observed in multiple campaigns the LC of each campaign was analyzed individually. Method: method that yielded best period selected by eye-inspection. Flag: reliability of determined periods. $\nu$: Flare frequency. We consider all periods greater than half the campaign duration as periods with less confidence. For periods larger than the campaign duration we do not give any uncertainties as they are the ones with the lowest confidence.}
\label{periodTable}
\\ \hline\hline
EPIC ID & Camp. & \multicolumn{2}{c}{$P_{\rm rot}$} & Method & Flag & \multicolumn{2}{c}{$R_{\mathrm{per}}$} & $S_{\mathrm{ph}}$ & $S_{\mathrm{flat}}$ & $N_{\rm Flares}$& $\nu$\\ 
&  & \multicolumn{2}{c}{[d]} &  &  & \multicolumn{2}{c}{[\%]} & [ppm] & [ppm] & & [$N_{\mathrm{flares}}/d$]\\\hline
\endfirsthead
\multicolumn{6}{l}{Table~\ref{periodTable} continued.}\\
\hline\hline
EPIC ID & Camp. & \multicolumn{2}{c}{$P_{\rm rot}$} & Method & Flag & \multicolumn{2}{c}{$R_{\mathrm{per}}$} & $S_{\mathrm{ph}}$ & $S_{\mathrm{flat}}$ & $N_{\rm Flares}$& $\nu$\\ 
&  & \multicolumn{2}{c}{[d]} &  &  & \multicolumn{2}{c}{[\%]} & [ppm] & [ppm] & & [$N_{\mathrm{flares}}/d$]\\\hline
\endhead
\hline
\multicolumn{6}{c}{to be continued on the next page}\\
\hline
\endfoot
\endlastfoot
202059183	&	C0	&	\multicolumn{2}{c}{---}			&	---	&	N	&	\multicolumn{2}{c}{---}			&	607.99	&	135.14	&	0	&	---	\\
202059188	&	C0	&	0.691	&	0.000	&	LS	&	Y	&	2.525	&	0.202	&	10520.91	&	2504.63	&	4	&	0.121	\\
202059190	&	C0	&	\multicolumn{2}{c}{---}			&	---	&	N	&	\multicolumn{2}{c}{---}			&	1012.63	&	93.06	&	0	&	---	\\
202059192	&	C0	&	35.220	&	1.169	&	SINE	&	?	&	0.468	&	0.051	&	1844.13	&	105.82	&	0	&	---	\\
202059193	&	C0	&	19.012	&	1.046	&	LS	&	?	&	0.403	&	0.071	&	1161.41	&	111.35	&	0	&	---	\\
202059195	&	C0	&	42.460	&		&	SINE	&	?	&	1.707	&	0.805	&	6447.70	&	161.18	&	0	&	---	\\
202059197	&	C0	&	\multicolumn{2}{c}{---}			&	---	&	N	&	\multicolumn{2}{c}{---}			&	1070.55	&	80.90	&	0	&	---	\\
202059198	&	C0	&	27.307	&	0.465	&	LS	&	?	&	0.830	&	0.029	&	2802.33	&	80.49	&	1	&	0.030	\\
\end{longtable}

\section{Multiplicity of the stars in the \textit{K2} M-dwarf sample}

\begin{longtable}[h]{lcccc}
\caption{Multiplicity status for our targets. We list here visual as well as close eclipsing and spectroscopic binaries. In total, we found 96 possible binaries or multiple systems.}
\label{binaryTable}
\\ \hline\hline
EPIC ID & Camp. & Name & Component & Separation [arcsec] \\\hline
\endfirsthead
\multicolumn{5}{l}{Table~\ref{binaryTable} continued.}\\
\hline\hline
EPIC ID & Camp. & Name & Component & Separation [arcsec] \\\hline
\endhead
\hline
\multicolumn{5}{c}{to be continued on the next page}\\
\hline
\endfoot
\endlastfoot
202059183	&	C0	&	Gaia DR2 3423743492225744128	&		&	7.2	\\
202059188	&	C0	&	WDS 06102+2234	&		&	1.9	\\
202059198	&	C0	&	WDS 06215+1618	&		&	115.0	\\
202059199	&	C0	&	Gaia DR2 3369198983818682624	&		&	119.0	\\
202059221	&	C0	&	WDS 06401+2835	&		&	0.3	\\
201367065	&	C1	&	WDS 11293-0127	&		&	26.0	\\
201481218	&	C1	&	WDS 11132+0014	&		&	0.2	\\
201482319	&	C1	&	WDS 11477+0016	&		&	24.7	\\
\hline
\end{longtable}

\begin{figure*}
  \centering
  \includegraphics[width=0.32\textwidth,angle=270]{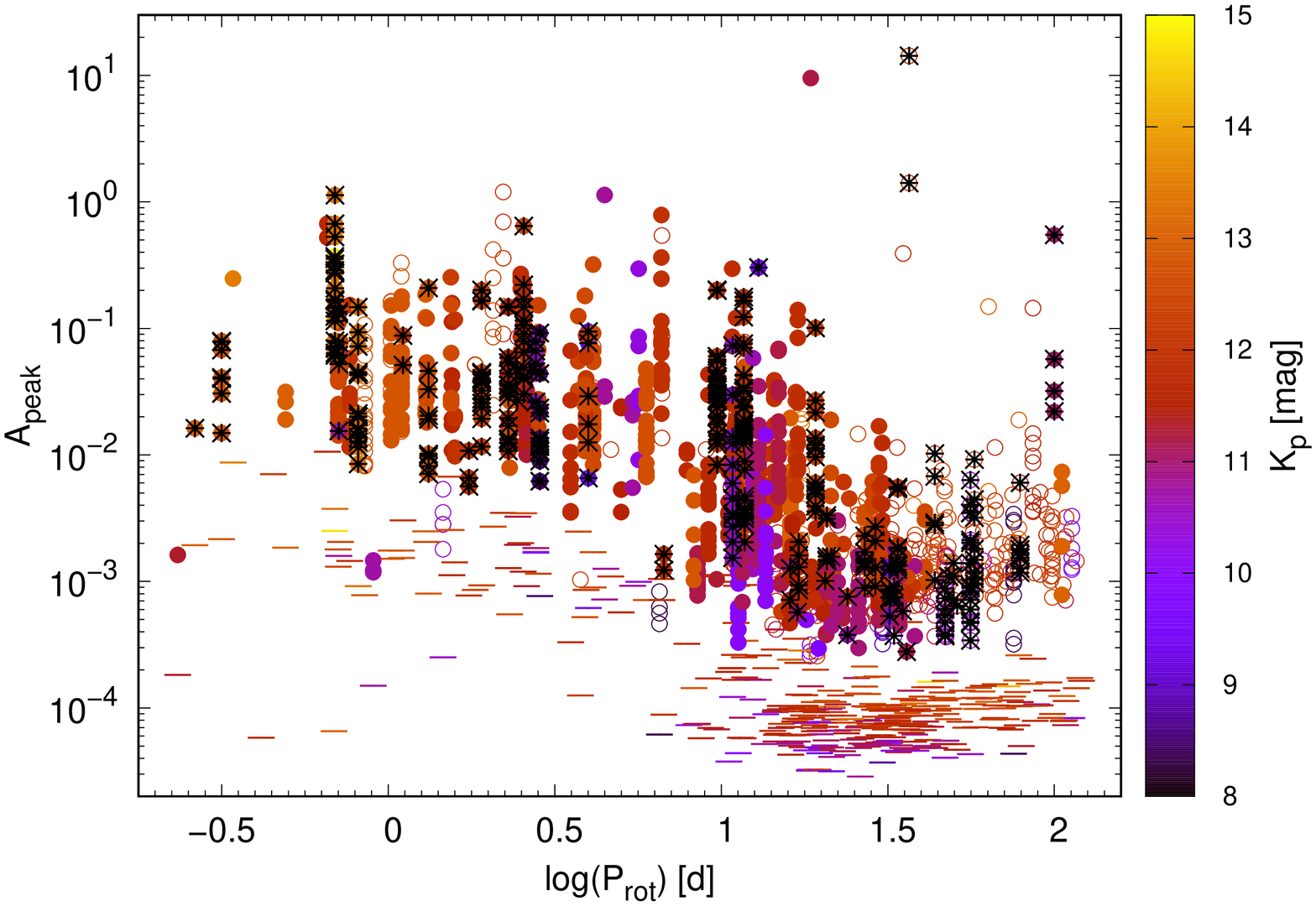}
  \includegraphics[width=0.32\textwidth,angle=270]{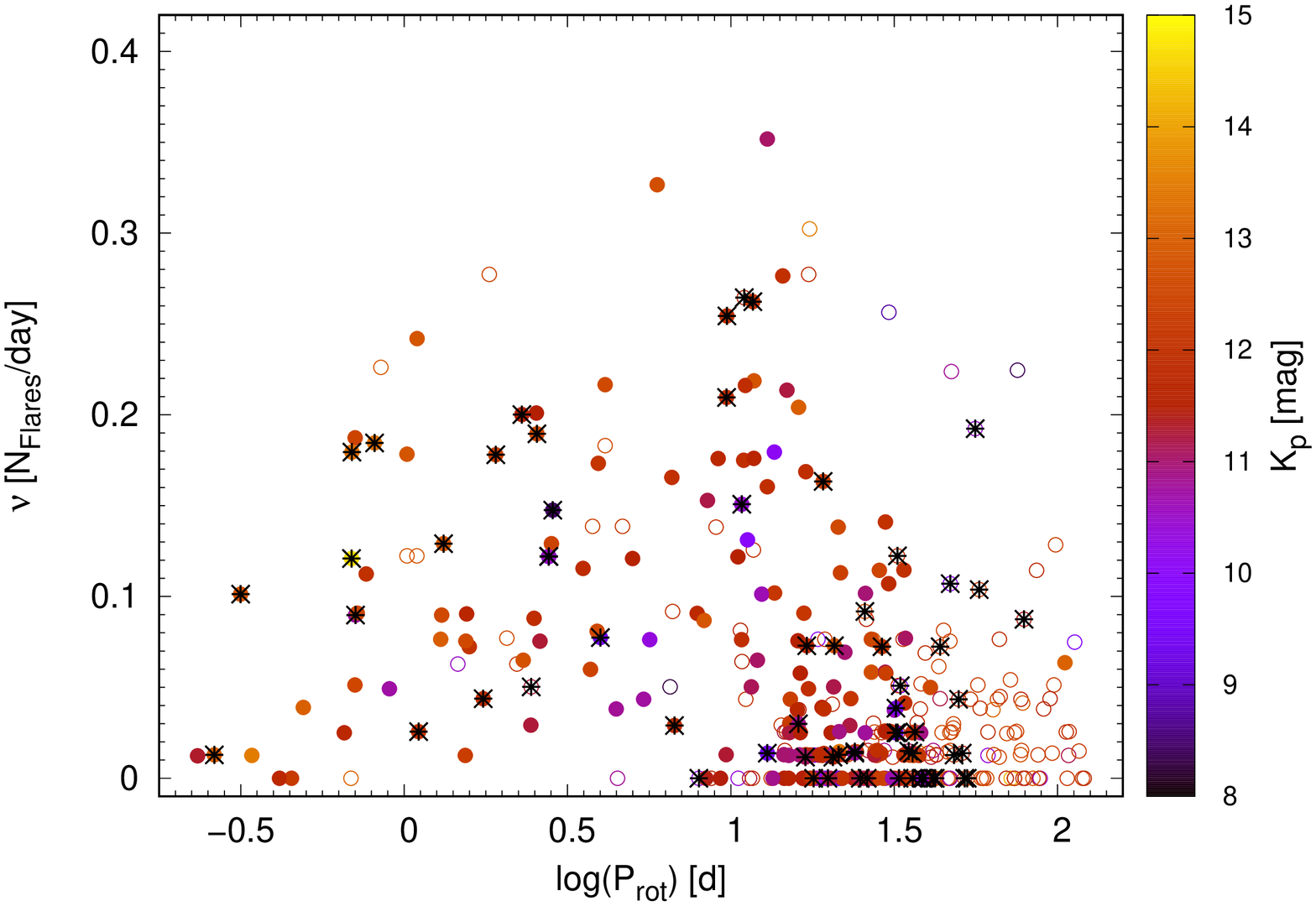}  
  \caption{Relation between rotation period and activity indicators related to flares for all 341 targets with measurable rotation period. Full and open symbols denote reliable periods and periods with less confidence, respectively. The black crosses mark the multiple systems that were explained in detail in Section~\ref{multiplicity}.}
  \label{period_activity_multiplicity}
\end{figure*}

\section{Example light curves}

\begin{figure*}
  \centering
  \includegraphics[width=0.7\textwidth,angle=270]{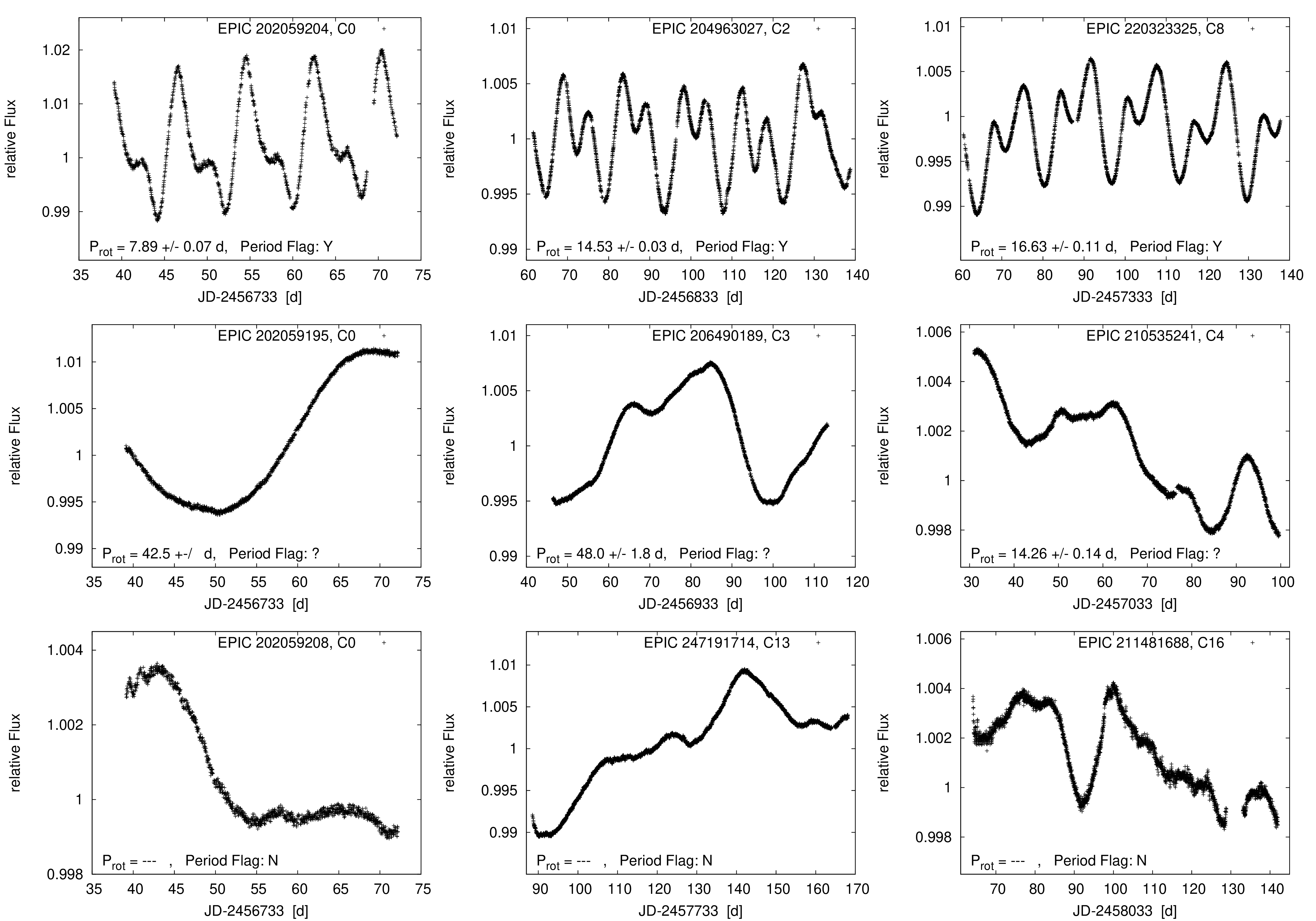}
  \caption{Various LCs that represent examples for the different period flags. The differnt rows show, from top to bottom, stars with an reliable period estimate (flag `Y'), stars with periods with lower confidence (flag `?') and stars with no detected period (flag `N'). }
  \label{Bsp_LC_K2}
\end{figure*}

\end{document}